\documentclass[%
 reprint,
superscriptaddress,
 amsmath,amssymb,
 aps,
]{revtex4-2}
\usepackage[normalem]{ulem}
\usepackage{comment}
\usepackage{graphicx}
\usepackage{algorithm}
\usepackage{algpseudocode}
\usepackage{amsmath}
\usepackage{dcolumn}
\usepackage{bm}
\usepackage{xcolor}
\definecolor{darkblue}{RGB}{0,40,180}

\begin{document}


\title{Free-Energy Differences from Nonequilibrium Fluctuations in High Dissipation}

\author{Sreekanth K Manikandan}
\email{sreekanth.manikandan@physics.gu.se}
\affiliation{Department of Physics, University of Gothenburg, 41296 Gothenburg, Sweden}%

\author{Giovanni Volpe}
\affiliation{Department of Physics, University of Gothenburg, 41296 Gothenburg, Sweden}%
\affiliation{Science for Life Lab, Department of Physics, University of Gothenburg, 41296 Gothenburg, Sweden}%

\date{\today}

\begin{abstract}
Equilibrium free-energy differences can be measured from nonequilibrium work fluctuations using fluctuation theorems, such as the Jarzynski equality and the Crooks fluctuation theorem.
However, these approaches become statistically inefficient in high-dissipation regimes because their convergence requires trajectories with negative entropy production, events that occur exponentially rarely. 
Here, we demonstrate that repeated measurements of trajectory fluctuations are sufficient to determine entropy production without relying on such rare events, or to determine a lower bound on it when only partial measurements are available.
This information then yields exact estimates of free-energy differences, or rigorous bounds.
We validate this approach, named EquiNET, with numerical simulations, including one of biomolecular folding and unfolding, showing that it recovers accurate free-energy estimates even in regimes where conventional approaches fail.
\end{abstract}


\maketitle

\begin{figure*}
    \centering
    \includegraphics[width=\linewidth]{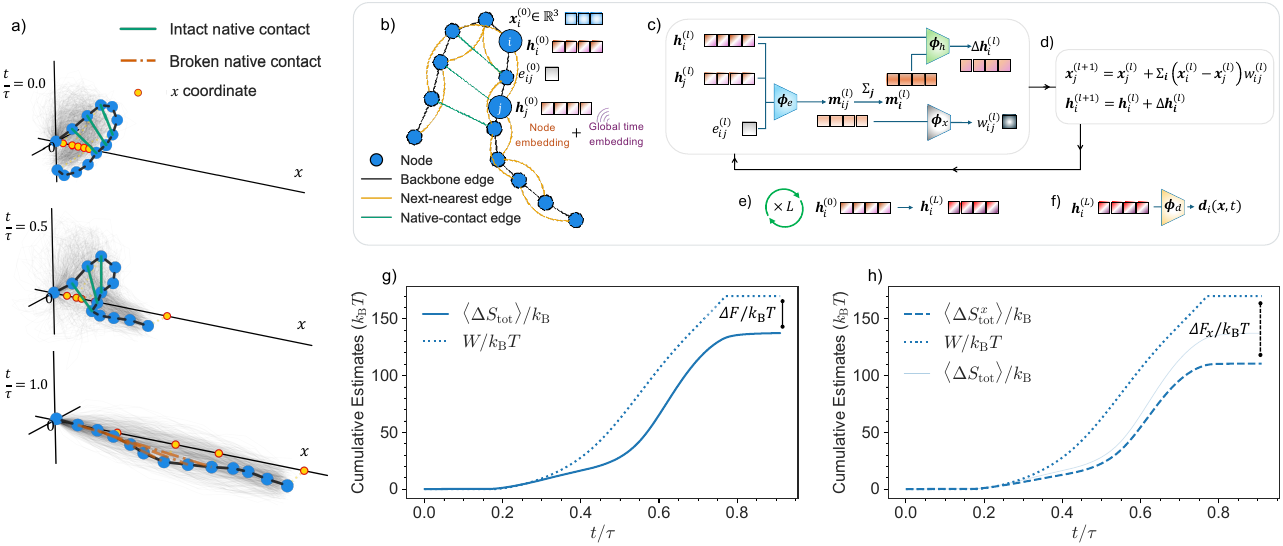}
    \caption{\textbf{EquiNET---Equilibrium free-energy differences from Non-Equilibrium Trajectories.}
    \textbf{(a)} Representative pulling trajectories of a biomolecule at normalized times $t/\tau=0$, $0.5$, and $1$, illustrating the unfolding process, where $\tau$ is the total duration of the pulling protocol.
    Polymer backbone (black solid line) with interaction sites (blue beads) along the molecule, intact native (green lines), and broken native contacts (orange dash-dotted lines).
    The $x$ coordinates of selected beads (yellow filled circles) define the coarse-grained representation.
    Gray curves in the background represent other realizations from the ensemble of trajectories generated by repeated applications of the same pulling protocol.
    \textbf{(b-f)} Time-conditioned E(3)-equivariant graph neural network (EGNN) for inference:
    \textbf{(b)} The polymer is represented as a graph whose nodes correspond to beads and whose edges encode backbone (black), next-nearest-neighbor (yellow), and native-contact interactions (green).
    Each node $i$ is associated with its Cartesian coordinates $\mathbf{x}_i^{(0)} \equiv \mathbf{x}_i^t\in\mathbb{R}^3$ and a feature vector $\mathbf{h}_i^{(0)}$ which is a sum of a learnable bead embedding and a global time embedding. 
    Each edge $(i,j)$ is assigned the scalar feature $e_{ij}^{(0)}$, corresponding to the Euclidean distance between the connected beads.
    \textbf{(c)} Edge messages $\mathbf{m}_{ij}^{(0)}$ are computed from the node and edge features using the multi-layer perceptron $\boldsymbol{\phi}_e$, and are aggregated to obtain the node message ${\bm m}_i^{(0)}$.
    The edge and node messages are then used to compute the coordinate-update weights $w_{ij}^{(0)}$ and the node-feature update $\Delta\mathbf{h}_i^{(0)}$, respectively.
    \textbf{(d)} EGNN update equations to obtain the updated node coordinates and node embeddings. 
    \textbf{(e)} The updated graph representation is propagated to the subsequent EGNN layers, and the message-passing and update operations are repeated for $L$ layers. 
    \textbf{(f)} The final node embeddings are passed through the readout network $\boldsymbol{\phi}_d$ to predict the bead-wise three-dimensional vector field $\mathbf{d}_i(\mathbf{x},t)$. This vector field is used to construct the variational objective in Eq.~\eqref{eq:tur} using trajectory data. The network parameters $\boldsymbol{\theta}$ are learned by maximizing the objective in Eq.~\eqref{eq:tur}, yielding an estimate of the instantaneous entropy production rate $\sigma(t)$.
    \textbf{(g)} When all the degrees of freedom of the polymer are tracked, integrating the inferred entropy production rate over the duration of the process yields the total entropy production, $\langle \Delta S_{\mathrm{tot}} \rangle$. Combining this estimate with the measured work according to $\Delta F=\langle W\rangle-T\langle\Delta S_{\mathrm{tot}}\rangle$ yields the equilibrium free-energy difference. The vertical separation between the work curve and the inferred entropy-production curve at the end of the protocol ($t/\tau=1$) corresponds to the estimated free-energy difference, $\Delta F/k_{\mathrm B}T$. 
    \textbf{(h)} When only the $x$ coordinates of few beads are measured, integrating the inferred entropy production rate over the protocol yields a lower bound $\langle \Delta S^x_{\mathrm{tot}} \rangle$ on the total entropy production, and a corresponding bound on the equilibrium free-energy difference given by $\Delta F_{x}/k_{\mathrm B}T$.}
    \label{fig:1}
\end{figure*}

The equilibrium free-energy landscape governs the stability, dynamics, and function of physical, chemical, and biological systems in contact with a heat bath \cite{landau2013statistical, chandler1987introduction}.
In biomolecules, for example, it determines the relative stability of folded and unfolded conformations \cite{dill1997levinthal}, the locations of transition states \cite{wolynes1995navigating, hanggi1990reaction}, and the kinetics of folding and conformational rearrangements \cite{frauenfelder1991energy, onuchic1997theory, dill2012protein,bustamante2026proteins}.
Determining it, however, generally requires extensive sampling of the equilibrium ensemble, which is a challenging task for complex systems with rugged energy landscapes and large free-energy barriers \cite{kollman1993free, valsson2016enhancing}. 

Nonequilibrium approaches offer an attractive alternative by reconstructing equilibrium free-energy differences from repeated driven transitions between equilibrium states \cite{jarzynski2011equalities}, thereby avoiding the need for direct equilibrium sampling.
Two central results underlying these approaches are the Jarzynski equality \cite{jarzynski1997nonequilibrium, Jarzynski},
\begin{equation}
\label{eq:JE}
\left\langle e^{-\beta W} \right\rangle = e^{-\beta \Delta F},
\end{equation}
which relates the exponential average of the work $W$ performed during a nonequilibrium process to the equilibrium free-energy difference $\Delta F$, and the Crooks fluctuation theorem \cite{crooks},
\begin{equation}
\label{eq:CFT}
P_{\mathrm{F}}(W)=P_{\mathrm{R}}(-W)e^{\beta(W-\Delta F)},
\end{equation}
where $P_{\mathrm{F}}(W)$ and $P_{\mathrm{R}}(W)$ are the work distributions for the forward and corresponding time-reversed processes, and  $\beta=(k_{\rm B}T)^{-1}$, with $k_{\rm B}$ the Boltzmann constant and $T$ the absolute temperature.
Equation~\eqref{eq:CFT} implies that the forward and reverse work distributions should intersect at \(W=\Delta F\), and therefore, the equilibrium free-energy difference can be obtained directly from their crossing point, given enough statistics.
The Second Law of Thermodynamics follows from Eq.~\eqref{eq:JE} through the application of Jensen's inequality \cite{cover1999elements} as the positivity of the average entropy production $\langle \Delta S_{\rm tot} \rangle = \beta\bigl(\langle W\rangle-\Delta F\bigr)\ge0$. 
Since their discovery, these results have been extensively validated in experiments ranging from mechanically manipulated biomolecules \cite{liphardt2002equilibrium, collin2005verification} to colloidal particles \cite{blickle2005thermodynamics} and electrical circuits \cite{garnier2005nonequilibrium}.
Together with the earlier versions of fluctuation theorems \cite{bochkov1981nonlinear,evans1993probability,gallavotti1995dynamical,gallavotti1995dynamical2}, these results marked major milestones in extending classical statistical physics and thermodynamics beyond the linear-response regime.
These developments established the modern framework of stochastic thermodynamics  \cite{Seifert:2012stf,Ciliberto:2017est,seifert2025stochastic} and have since inspired numerous theoretical generalizations across a broad range of classical \cite{seifert2005entropy,sagawa2010generalized,argun2016non,camunas:ebe} and quantum \cite{mukamel2003quantum,talkner2007fluctuation,campisi2011colloquium} nonequilibrium systems.

Despite this exact and general significance, both identities share a well-recognized practical limitation: their convergence relies on trajectories with negative entropy production, which occur only exponentially rarely \cite{crooks, gallavotti1995dynamical}.
For the Jarzynski equality, typical trajectories satisfy $W>\Delta F$, so $e^{-\beta(W-\Delta F)}<1$; recovering the equality in Eq.~\eqref{eq:JE} therefore requires contributions from the rare trajectories with $W<\Delta F$.
Similarly, for the Crooks fluctuation theorem, $\Delta F$ is determined by the crossing point between the forward and reverse work distributions, which also requires sampling rare trajectories.
As a result, in moderate-to-strongly dissipative regimes ($\langle \Delta S_{\rm tot}\rangle$ larger than a few tens of $k_{\rm B}T$), both identities require impractically large numbers of nonequilibrium realizations to converge \cite{collin2005verification, gore2003bias}.

The underlying challenge is that $\Delta S_{\rm tot}$, which comprises both the system entropy change and the entropy transferred to the environment through heat dissipation, is generally not directly accessible in experiments in the same way as the work. 
In single-molecule experiments, for instance, there is neither a direct calorimetric probe available for the dissipated heat \cite{wang2020nanocalorimeters} nor experimental access to the underlying molecular energy landscape \cite{henzler2007dynamic}.  
Similarly, the system entropy contribution requires complete knowledge of the microscopic state probabilities, which is not directly measurable.
This leads to the following question: can one obtain an independent estimate of $\langle \Delta S_{\mathrm{tot}}\rangle$, which can then be exploited to estimate equilibrium free-energy differences?

In this work, we demonstrate that this is possible.
We introduce EquiNET, a data-driven technique for inferring equilibrium free-energy differences from nonequilibrium  trajectory fluctuations.
EquiNET builds on recent trajectory-based inference techniques that infer $\langle \Delta S_{\mathrm{tot}}\rangle$ (or a lower bound to it) independently of fluctuation theorems and use it to correct for dissipative losses to yield $\Delta F$ (or rigorous bounds on it). 
Importantly, we demonstrate that EquiNET remains effective in strongly dissipative regimes where conventional fluctuation-theorem-based estimators fail to converge.

Our approach relies on a variational optimization procedure derived from the short-time thermodynamic uncertainty relation (TUR) \cite{manikandan2020inferring, van2020entropy, Shun:eem, otsubo2022estimating, manikandan2024estimate, manikandan2021quantitative, das2026localising}, which expresses the instantaneous entropy production rate, $\sigma(t)$,  in overdamped systems, as the solution of the optimization problem,
\begin{equation}
\label{eq:tur}
\sigma(t)=\frac{1}{\mathrm{d}t}\max_{\bm d}
\left[
\frac{2k_{\rm B}\langle J_{\bm d}\rangle^2}
{\mathrm{Var}(J_{\bm d})}
\right],
\end{equation}
where $J_{\bm d}=\bm d(\bm x(t),t)\circ\mathrm{d}\bm x(t)$ is a weighted stochastic current defined through the Stratonovich product, and the averages are taken over the joint distribution $p(\bm x(t),\bm x(t+\mathrm{d}t))$. The optimization is performed over the coefficient field $\bm d(\bm x,t)$, whose optimum satisfies $\bm d^*(\bm x,t)\propto\bm{\mathcal F}(\bm x,t)$, where $\bm{\mathcal F}$ denotes the thermodynamic force field \cite{manikandan2020inferring,Shun:eem,van2020entropy, otsubo2022estimating, das2026localising}. The key insight underlying this representation is that, over sufficiently short times, the accumulated entropy production remains small, so short trajectory segments with negative entropy production occur with a much higher relative likelihood than longer trajectories segments. This intuition is supported by rigorous results: for overdamped diffusive processes, entropy production fluctuations have been shown to become progressively Gaussian \cite{manikandan2020inferring, van2020entropy, Shun:eem, otsubo2022estimating} and less positively skewed \cite{manikandan2022nonmonotonic} in the short-time limit. Eq.~\eqref{eq:tur} exploits this asymptotic behavior to infer entropy production using only the first two cumulants of $J_{\bm d}$, thereby explicitly avoiding higher order statistics and exponential averages (see Ref.~\cite{otsubo2022estimating} for a detailed discussion and proof).
Since the inference relies solely on the observed trajectory fluctuations, the method naturally extends to situations in which only a subset of the system's degrees of freedom is accessible, yielding a lower bound on the entropy production and, consequently, bounds on the equilibrium free-energy difference.

We consider a \(d\)-dimensional overdamped system with configuration vector \(\bm{x}(t) \in \mathbb{R}^d\), evolving in contact with a thermal reservoir at temperature \(T\).
The system is initially prepared in an equilibrium state \(A\) and is driven to a final equilibrium state \(B\) by a time-dependent protocol \(\lambda(t)\). 
The protocol of total duration $\tau$ consists of three stages: 
an initial equilibration plateau, during which the control parameter is held fixed at its initial value, $\lambda(t)=\lambda_0\equiv\lambda(0)$;
a finite-time driving ramp, during which $\lambda(t)$ is varied continuously from $\lambda_0$ to its final value $\lambda_\tau\equiv\lambda(\tau)$; 
and a final equilibration plateau, during which $\lambda(t)=\lambda_\tau$ is held constant. 
The durations of the initial and final equilibration plateaus are chosen based on the characteristic relaxation times of the system, allowing it to relax (sufficiently close) to equilibrium.
As a general guideline, the equilibration plateau should be at least as long as the longest relevant relaxation time of the system, and preferably several times longer.
In practice, as we discuss later, both the adequacy of the plateau duration and the limitations of assuming equilibration within a finite observation window can be assessed empirically.

As a representative example that we revisit later, Fig.~\ref{fig:1}(a) shows an ensemble of trajectories generated during repeated force-induced unfolding experiments on a polymer hairpin. 
The polymer is initially prepared in a folded configuration stabilized by native contacts between specific beads.
As the pulling protocol progresses, these contacts progressively break, causing the molecule to unfold and explore a heterogeneous ensemble of nonequilibrium trajectories before reaching the fully extended, unfolded state.

We assume that the stochastic dynamics of the system can be described using an overdamped Langevin equation,
\begin{equation}
\label{eq:langevin}
\dot{\bm{x}}(t) = \boldsymbol{\mu} \cdot \mathbf{F}(\bm{x}(t),\lambda(t)) + \boldsymbol{\xi}(t),
\end{equation}
where $\boldsymbol{\mu}$ is the mobility tensor, $\mathbf{F}(\bm{x},\lambda) = -\nabla_{\bm{x}} U(\bm{x},\lambda)$ is the force derived from a time-dependent potential $U(\bm{x},\lambda)$, and $\boldsymbol{\xi}(t)$ is a Gaussian white noise satisfying
\begin{equation}
\langle \boldsymbol{\xi}(t) \rangle = 0, \qquad 
\langle \boldsymbol{\xi}(t)\boldsymbol{\xi}^\top(t') \rangle = 2 k_{\rm B} T \, \boldsymbol{\mu} \, \delta(t - t').
\end{equation}
Each realization of the dynamics generates a trajectory $\{\bm{x}(t)\}$, along which a stochastic work $W[\bm{x}(t)] = \int_{0}^\tau dt\;  \partial_\lambda U \dot{\lambda}$ is performed by the external driving $\lambda(t)$ \cite{seifert2008stochastic}. Assuming that the process begins and ends at equilibrium states, the total entropy production associated with this process can be identified as \cite{seifert2005entropy}
\begin{align}
\begin{split}
\frac{\Delta S_{\mathrm{tot}}}{k_{\rm B}} &= -\beta Q +\frac{\Delta S_{\rm sys}}{k_{\rm B}}\\ &=-\beta (\Delta U - W) +\frac{\Delta S_{\rm sys}}{k_{\rm B}}\\
&=\beta W - \beta (\Delta U - T\Delta S_{\rm sys}),    
\end{split}
\end{align}
where $Q[\bm{x}(t)]=\int_0^\tau \bm{F}(\bm{x}(t),\lambda(t))\circ d\bm{x}(t)$ is the heat absorbed by the system from the thermal environment, and the symbol ``$\circ$'' denotes the Stratonovich integral (i.e., the force is evaluated at the midpoint of each infinitesimal displacement). The change in internal energy is $\Delta U=U(\bm{x}_\tau,\lambda_\tau)-U(\bm{x}_0,\lambda_0)$, such that the First Law of Thermodynamics reads $\Delta U=Q+W$. The change in stochastic system entropy is $\Delta S_{\rm sys}=-k_{\rm B}\ln[p(\bm{x}_\tau)/p(\bm{x}_0)]$, where $p(\bm{x}_t)$ denotes the probability density of finding the system in configuration $\bm{x}$ at time $t$.
Upon ensemble averaging, we get
\begin{equation}
    \label{ep}
    \langle \Delta S_{\rm tot} \rangle /k_{\rm B} = \beta (\langle W \rangle  - \Delta F),
\end{equation}
where $\Delta F = \langle \Delta U \rangle  - T\langle \Delta S_{\rm sys} \rangle = F_B - F_A$ is the Helmholtz free energy difference between the initial and final equilibrium states, and the quantity $\langle W \rangle  - \Delta F$ is the dissipated work. 
If the end state has not reached equilibrium (or a state sufficiently close to it) within the observation window $[0,\tau]$, the measured entropy production provides a lower bound on the dissipated work \cite{seifert2005entropy}.

EquiNET estimates $\Delta F$ by inferring the total entropy production $\langle \Delta S_{\mathrm{tot}} \rangle$ from trajectory fluctuations.
Specifically, it uses Eq.~\eqref{eq:tur} to perform a variational inference of the time-dependent entropy production rate $\sigma(t)$ on short trajectory segments and then integrates these values over the entire duration of the process to obtain $\langle \Delta S_{\mathrm{tot}} \rangle$.
Since the estimator in Eq.~\eqref{eq:tur} depends only on the mean and variance of short trajectory currents at a given time $t$  \cite{manikandan2020inferring, Shun:eem, van2020entropy, otsubo2022estimating, manikandan2022nonmonotonic}, it is expected to converge without relying on long trajectories having accumulated rare, $\Delta S_{\rm tot} < 0$ events, as required by the fluctuation theorems.

The variational representation of $\sigma(t)$ in Eq.~\eqref{eq:tur} suggests a data-driven optimization approach: 
we approximate the unknown dissipative force field $\bm{\mathcal F}(\bm x,t)$ using a parameterized ansatz ${\bm d}(\bm x,t\vert\bm\theta)$ and learn $\bm\theta$ from data \cite{manikandan2020inferring, Shun:eem,otsubo2022estimating}.
Interestingly, when a suitable basis representation of $\bm{\mathcal F}(\bm x,t)$ is available, the resulting optimization admits an exact analytical solution \cite{van2020entropy, das2022inferring}; we will return to this analytically tractable case shortly.
In generic cases, however, no such basis is known \emph{a priori} and the optimal dissipative force field may depend on complex, many-body interactions.
This motivates the use of expressive function approximators that can learn the force field directly from trajectory data. At the same time, incorporating the underlying physical symmetries and invariances into the model architecture can substantially simplify the inference problem and improve data efficiency \cite{karniadakis2021physics, yan2022physics}.
With these considerations in mind, we parameterize the dissipative force field ${\bm d}(\bm x,t)$ for the polymer hairpin using an equivariant graph neural network \cite{satorras2021n} operating on the molecular graph shown schematically in Fig.~\ref{fig:1}(b), where beads are represented as nodes and pairwise interactions define the graph connectivity.
Each node $i$ is associated with its Cartesian coordinates $\mathbf{x}_i^{(0)} \equiv \mathbf{x}_i^t\in\mathbb{R}^3$ and a feature vector $\mathbf{h}_i^{(0)}$ which is a sum of a learnable bead embedding and a global time embedding. 
Each edge $(i,j)$ is assigned a scalar feature $e_{ij}^{(0)}$, corresponding to the Euclidean distance between the beads.
The network employs an equivariant message-passing architecture, illustrated in Figs.~\ref{fig:1}(c,d), that captures many-body interactions while preserving rotational, translational, and permutation symmetries.
The resulting node embeddings (Fig.~\ref{fig:1}(e)) are then mapped by the readout network to an approximation of the dissipative force field $\bm{d}(\bm{x},t)$, conditional on the network parameters $\bm \theta$.
These parameters are then optimized end-to-end by maximizing the variational objective in Eq.~\eqref{eq:tur} to obtain the instantaneous entropy production rate $\sigma(t)$. 
The final step (Fig.~\ref{fig:1}(f)) then consists of integrating the instantaneous entropy production rate $\sigma(t)$ over the duration of the protocol to obtain the total entropy production $\langle \Delta S_{\rm tot} \rangle$, which is then combined with the measured average work  to infer the equilibrium free-energy difference $\Delta F$ in Fig.~\ref{fig:1}(g). 
Appendix A.1 provides details of the employed neural network and A.2 provides the details of this inference procedure.

The same procedure naturally extends to partial observations in which only a subset of the system's degrees of freedom is accessible. For the polymer hairpin, this could, for instance, correspond to tracking only the $x$ coordinates of the terminal bead and the beads that participate in the native contacts. In this case, inferred entropy production provides a rigorous lower bound, $\langle \Delta S_{\rm tot}^{x} \rangle$, on the true entropy production \cite{seifert2019stochastic, lanza2025measuring}, yielding a corresponding rigorous bound, $\Delta F^{ x}$, on the equilibrium free-energy difference. This is shown in Fig.~\ref{fig:1}(h).
Appendix A.3 provides the details of the coarse-grained inference procedure.

\begin{figure}
    \centering
    \includegraphics[width=\linewidth]{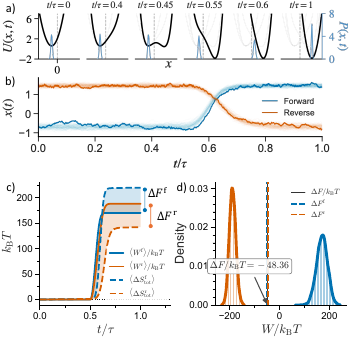}
    \caption{
    \textbf{Free-energy estimation with EquiNET for a driven bistable system.}
    \textbf{(a)} Snapshots of the time-dependent potential (black) and corresponding probability density (blue) during the forward protocol.
    The particle is initially confined in the left harmonic well, traverses an asymmetric double-well during the driving stage, and finally relaxes into the right harmonic well.
    \textbf{(b)} Representative forward (blue) and reverse (orange) trajectories illustrating the nonequilibrium transport between the two metastable basins. 
    A single trajectory in each direction is highlighted. Curves in the background represent  other realizations from the ensemble of trajectories generated by repeated applications of the same protocol.
    \textbf{(c)} Accumulated work (solid) and inferred total entropy production (dashed) for the forward and reverse protocols.
    Their difference, \(\Delta F=\langle W\rangle-T\Delta S_{\mathrm{tot}}\), converges to the equilibrium free-energy difference despite the strongly irreversible driving.
    \textbf{(d)} EquiNET accurately recovers the free-energy difference even though the forward (blue) and reverse (orange) work distributions do not overlap, making fluctuation-relation approaches based on rare trajectory sampling impractical for estimating $\Delta F$.}
    \label{fig:2}
\end{figure}

In the following, we first demonstrate EquiNET in an analytically tractable case, for a one-dimensional overdamped Langevin particle driven between two equilibrium states by a time-dependent potential \(U(x,t)\). The protocol continuously transforms the potential from an initial harmonic well into an asymmetric double-well and subsequently into another harmonic well centered at a different minimum, as illustrated in Fig.~\ref{fig:2}(a). This is achieved by interpolating the potential according to
\begin{equation}
U(x,t)
=
\alpha(t)U_L(x)
+\beta(t)U_{\rm DW}(x)
+\gamma(t)U_R(x),
\end{equation}
where the weights satisfy \(\alpha+\beta+\gamma=1\) and are chosen to be continuous functions whose first derivatives are also continuous. Their explicit functional forms and further details of the model are provided in Appendix~B. The intermediate potential is given by
\begin{equation}
U_{\rm DW}(x)=(x^2-1)^2+ax^3,
\end{equation}
whose cubic term of strength $a$  introduces an asymmetry between the two minima. Denoting the minima by \(x_L\) and \(x_R\), we define the local harmonic approximations
\begin{equation}
U_{L,R}(x)
=
U_{\rm DW}(x_{L,R})
+\frac12k_{L,R}(x-x_{L,R})^2,
\end{equation}
where \(k_{L,R}=\partial_x^2U_{\rm DW}(x_{L,R})\). The asymmetry produces a finite equilibrium free-energy difference between the two basins. 

Under the forward protocol, the particle is driven from the equilibrium state localized in the left basin to that localized in the right basin, as illustrated by the evolving potential and probability density in Fig.~\ref{fig:2}(a). The reverse protocol drives the system in the opposite direction under the time-reversed potential, \(U^{\rm rev}(x,t)=U(x,\tau-t)\). Representative trajectories for both protocols are shown in Fig.~\ref{fig:2}(b), with one trajectory highlighted by the solid line in each direction.
Curves in the background represent  other realizations from the ensemble of trajectories generated by repeated applications of the same protocol.

For this one-dimensional system, the variational optimization in Eq.~\eqref{eq:tur} can be solved analytically by expanding $d(x,t)$ in a finite polynomial basis \cite{van2020entropy, das2022inferring}.
Following Ref.~\cite{van2020entropy}, we represent \(d(x,t)\) as \(d(x,t)=\sum_{i=0}^{3}\theta_i(t)x^i\). Substituting this expansion into the definition of the stochastic current \(J_d\) in Eq.~\eqref{eq:tur} gives \(J_d(t)=\sum_{i=0}^{3}\theta_i(t)J_i(t)\), where \(J_i(t)=x^i(t)\circ\mathrm{d}x(t)\) are the corresponding basis currents.
From an ensemble of short-time trajectory increments, we estimate the mean current vector \(\langle\mathbf J(t)\rangle\) and covariance matrix \(\mathbf C(t)\), with elements \(C_{ij}(t)=\langle J_i(t)J_j(t)\rangle-\langle J_i(t)\rangle\langle J_j(t)\rangle\). Ref.~\cite{van2020entropy} showed that, for such a finite basis expansion, the optimization problem in Eq.~\eqref{eq:tur} admits an analytical solution, which yields the instantaneous entropy production rate as \(\sigma(t)=2k_{\rm B}\langle\mathbf J(t)\rangle^\top\mathbf C^{-1}(t)\langle\mathbf J(t)\rangle/\mathrm{d}t\).

The results of applying this inference procedure to the simulated trajectories are summarized in Figs.~\ref{fig:2}(c,d). 
The solid curves in Fig.~\ref{fig:2}(c)  denote the accumulated work and the dashed curves the inferred total entropy production. Although both quantities increase substantially during the nonequilibrium transition, their difference, \(\Delta F=\langle W\rangle-T\langle \Delta S_{\rm tot} \rangle\), converges to the same value, which agrees with the analytically computed equilibrium free-energy difference $\Delta F= -48.36\;k_{\rm B}T$ as shown in Figs.~\ref{fig:2}(d). See Appendix B for details of the calculation. Remarkably, this agreement is achieved even though the corresponding forward and reverse work distributions, shown in Fig.~\ref{fig:2}(d), are well separated and exhibit essentially no overlap, because the rare trajectories required for standard fluctuation-theorem-based free-energy estimation are absent from the sampled ensembles, making such estimates impractical with the available data. In contrast, the present inference recovers the equilibrium free-energy difference from either driving direction without requiring overlap of the work distributions.

\begin{figure*}
    \centering
    \includegraphics[width=1.0\linewidth]{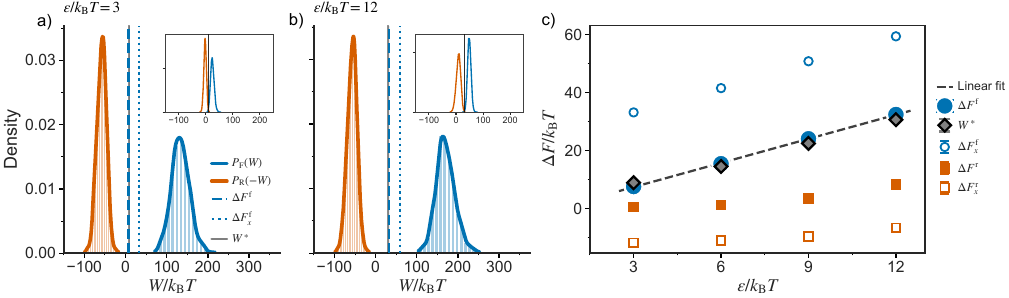}
    \caption{
    \textbf{EquiNET equilibrium free-energy differences for the polymer hairpin model.}
    \textbf{(a-b)} Forward ($P_{\mathrm{F}}(W)$, blue) and reverse ($P_{\mathrm{R}}(-W)$, orange) work distributions for native-contact strengths
    \textbf{(a)} $\epsilon/k_{\mathrm{B}}T=3$ and \textbf{(b)} $\epsilon/k_{\mathrm{B}}T=12$.
    The main panels show the work distributions obtained from the nonequilibrium pulling protocol used for inference, with dimensionless pulling velocity $\tilde{v}_x=2$. The insets show the corresponding distributions obtained using a slower-driving protocol with $\tilde{v}_x=0.2$, which lie closer to one another, and exhibit an overlap.
    The thick blue dashed lines indicate the free-energy differences $\Delta F^{\rm f}$ inferred with EquiNET using the full bead trajectories in the forward process, while the blue dotted lines show the corresponding estimates $\Delta F^{\rm f}_{x}$ obtained using only the $x$ coordinates of selected beads in the forward process. The gray vertical lines mark the crossing points $W^\ast$ of the forward and reverse work distributions under slow driving (inset), providing independent reference estimates of the equilibrium free-energy differences. As $\epsilon$ increases, the forward and reverse work distributions of the nonequilibrium protocol become increasingly separated; nevertheless, EquiNET recovers free-energy differences in good agreement with $W^\ast$.
    \textbf{(c)} Equilibrium free-energy difference as a function of $\epsilon$.
    Filled symbols denote estimates obtained from the full bead trajectories, while open symbols denote estimates obtained using only the $x$ coordinates of selected beads. Blue circles and orange squares indicate estimates from the forward and reverse protocols, respectively, while red diamonds show $W^\ast$ obtained from the slow-driving work distributions. The dashed line is a linear fit, to show the linear dependence of $\Delta F$ on $\epsilon$.
    Error bars indicate the standard deviation over six independent runs. They are smaller than the marker size and are therefore not visible in the plots; the corresponding numerical values are reported in Appendix~\ref{app:free_energy_estimates}.
    }
    \label{fig:3}
\end{figure*}

As a second example, we consider the polymer hairpin model introduced in Fig.~\ref{fig:1}(a).
The system consists of $n=13$ beads with three-dimensional
coordinates $\{\mathbf r_i\}_{i=1}^{n}$. 
The total potential energy is written as 
$U(\mathbf r,t)
=
U_{\mathrm{bond}}
+
U_{\mathrm{bend}}
+
U_{\mathrm{native}}
+
U_{\mathrm{rep}}
+
U_{\mathrm{anchor}}
+
U_{\mathrm{pull}}(t)$.
Adjacent beads are connected by stiff harmonic bonds,
\begin{equation}
U_{\mathrm{bond}}
=
\frac{k_{\mathrm{bond}}}{2}
\sum_{i=1}^{n-1}
\left(
|\mathbf r_{i+1}-\mathbf r_i|-b_i
\right)^2,
\end{equation}
where \(b_i\) denotes the equilibrium length of the \(i\)-th bond.
Chain stiffness is introduced through the discrete bending energy
\begin{equation}
U_{\mathrm{bend}}
=
\frac{k_{\mathrm{bend}}}{2b_{\mathrm{ref}}^2}
\sum_{i=2}^{n-1}
\left|
\mathbf r_{i+1}
-2\mathbf r_i
+\mathbf r_{i-1}
\right|^2,
\end{equation}
where $
b_{\mathrm{ref}}
=
\frac{1}{n-1}
\sum_{i=1}^{n-1} b_i$ 
is the mean equilibrium bond length.

To stabilize the folded hairpin configuration, selected non-neighboring
bead pairs are assigned attractive Lennard--Jones interactions,
\begin{equation}
\label{eq:native}
U_{\mathrm{native}}
=
\sum_{(i,j)\in\mathcal N}
4\epsilon_{\mathrm{native}}
\left[
\left(\frac{\sigma_{ij}}{r_{ij}}\right)^{12}
-
\left(\frac{\sigma_{ij}}{r_{ij}}\right)^6
\right],
\end{equation}
where
\(r_{ij}=|\mathbf r_i-\mathbf r_j|\), and
\(\mathcal N=\{(4,7),(3,8),(2,9)\}\)
denotes the set of native contacts, and $\sigma_{ij}$ are the
corresponding interaction lengths. All remaining non-neighboring, non-native bead pairs interact through a
purely repulsive Weeks--Chandler--Andersen potential,
\begin{equation}
U_{\mathrm{rep}}(r)
=
\scalebox{0.7}{$\begin{cases}
4\epsilon_{\mathrm{rep}}
\left[
\left(\dfrac{\sigma_{\mathrm{rep}}}{r}\right)^{12}
-
\left(\dfrac{\sigma_{\mathrm{rep}}}{r}\right)^6
\right]
+\epsilon_{\mathrm{rep}},
&
r<2^{1/6}\sigma_{\mathrm{rep}},
\\[6pt]
0,
&
r\ge 2^{1/6}\sigma_{\mathrm{rep}}.
\end{cases}$}
\end{equation}
Parameter values of the model are provided in
Appendix~C. All quantities are expressed in reduced units, with $k_{\mathrm B}T$ as the
unit of energy and $\ell_0$ as the unit of length. The corresponding
diffusive time scale is
$t_0=\ell_0^2/(\mu k_{\mathrm B}T)$, where $\mu$ is the mobility.
Accordingly, velocities are expressed in units of $\ell_0/t_0$ and
forces in units of $k_{\mathrm B}T/\ell_0$. We set
$k_{\mathrm B}T=\ell_0=t_0=1$ in the simulations.

To mechanically drive the hairpin between its folded and extended states, we apply harmonic constraints to its terminal beads. The first bead is confined by a stiff harmonic anchor,
\begin{equation}
U_{\mathrm{anchor}}
=
\frac{k_{\mathrm{anchor}}}{2}
\left|
\mathbf r_1-\mathbf r_1^{\mathrm{trap}}
\right|^2,
\end{equation}
while the final bead is coupled to a time-dependent harmonic pulling trap,
\begin{equation}
U_{\mathrm{pull}}(t)
=
\frac{k_{\mathrm{pull}}}{2}
\left|
\mathbf r_n-\boldsymbol{\lambda}(t)
\right|^2.
\end{equation}
Each realization begins with an equilibration plateau during which the
pulling trap is held fixed at its initial position
\(\boldsymbol{\lambda}_0\). The forward driving stage subsequently translates the trap at a constant velocity $v_x$ along the $x$-direction. 
The corresponding dimensionless pulling velocity is $\tilde{v}_x\equiv v_x t_0/\ell_0$. 
After the trap reaches its final position,
it is held fixed during a final equilibration plateau. The reverse protocol
is obtained by time-reversing the complete forward trap trajectory,
\begin{equation}
\boldsymbol{\lambda}_{\mathrm{rev}}(t)
=
\boldsymbol{\lambda}_{\mathrm{fwd}}(\tau-t).
\end{equation}
The dynamics follows overdamped Langevin equation in Eq.~\eqref{eq:langevin} with $\mathbf{F}_i=-\nabla_{\mathbf{r}_i}U(\mathbf{r},t)$. 
During the pulling protocol, the hairpin progressively unfolds through the sequential rupture of native contacts, as illustrated in Fig.~\ref{fig:1}(a).
Work is performed only through displacement of the pulling trap. For a
trap displacement \(d\boldsymbol{\lambda}\),
\begin{equation}
\delta W
=
\frac{\partial U_{\mathrm{pull}}}
{\partial\boldsymbol{\lambda}}
\cdot
d\boldsymbol{\lambda}
=
-k_{\mathrm{pull}}
\left(
\mathbf r_n-\boldsymbol{\lambda}
\right)
\cdot
d\boldsymbol{\lambda}.
\end{equation}
In the numerical simulations, this stochastic work increment is evaluated
using the Stratonovich discretization, and is integrated over the duration of the protocol to obtain the accumulated work $W$.

\begin{figure}
    \centering
    \includegraphics[width=\linewidth]{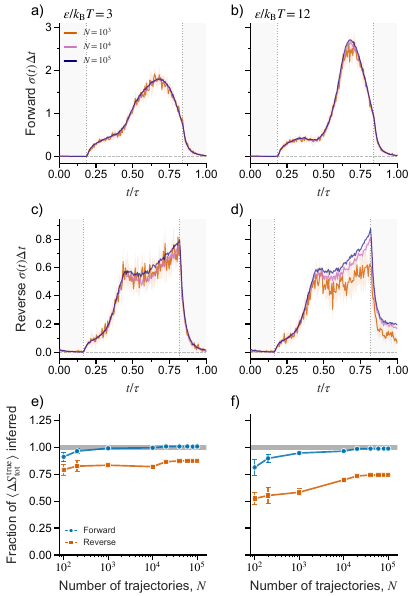}
    \caption{
    \textbf{Convergence of the EquiNET entropy-production estimates.}
    \textbf{(a,b)} Forward and \textbf{(c,d)} reverse entropy-production rates, $\sigma(t)\Delta t$, as functions of normalized time, $t/\tau$, for interaction strengths $\epsilon/k_{\mathrm B}T=3$ and $12$, respectively.
    For each $\epsilon$, the colored curves correspond to the total numbers of trajectories used for inference: $N=10^3$ (orange), $10^4$ (purple), and $10^5$ (blue).
    Solid curves show the mean over six independent runs, and the corresponding shaded bands indicate $\pm 1$ standard deviation.
    The $N=10^{5}$ estimate  serves as the large-sample reference.
    Gray shaded regions denote the initial and final equilibration periods, during which the pulling trap is held fixed.
    Increasing $N$ suppresses statistical fluctuations while preserving the overall temporal profile of the inferred entropy production. 
    The persistent entropy production during the final equilibration period in the reverse process indicates residual relaxation beyond the observation window.
    \textbf{(e,f)} Fraction of the true total entropy production inferred as a function of the number of trajectories used for training,
    $\langle \Delta S_{\mathrm{tot}}(N)\rangle/\langle \Delta S_{\mathrm{tot}}^{\mathrm{true}}\rangle$ for $\epsilon/k_{\mathrm B}T=3$ and $12$.
    Blue circles and orange squares denote the forward and reverse  estimates, respectively, and error bars indicate the standard deviation
    over six independent runs. 
    The reverse estimates systematically lie below  $\langle \Delta S_{\mathrm{tot}}^{\mathrm{true}}\rangle$ because the system has not fully relaxed by the end of the reverse protocol.
    }
    \label{fig:4}
\end{figure}

Unlike in the one-dimensional example in Fig.~\ref{fig:2}, where ${\bm d}({\bm x}, t)$ can be represented using a small finite set of polynomial basis functions, extending this explicit basis expansion to the polymer hairpin becomes computationally intractable.
The thermodynamic force in this case is a high-dimensional vector field defined over the full configuration space of the hairpin, and a comparable multivariate basis expansion would require a rapidly growing number of basis currents as the non-linearity and the number of degrees of freedom increase.
Consequently, the analytical optimization procedure we used in the one-dimensional case does not scale efficiently to this system.
To overcome this limitation, we parameterize the coefficient field using a time-conditioned, E(3)-equivariant graph neural network (EGNN \cite{satorras2021n}), as introduced in Figs.~\ref{fig:1}(b--f). The network takes the bead coordinates \(\{\mathbf r_i\}_{i=1}^{n}\) and time $t$ as input and returns a vector-valued coefficient field acting on every bead ${\bm d}(\bm r,t)$. The network parameters are then optimized end-to-end according to the variational objective in Eq.~\eqref{eq:tur}, to infer entropy production directly from the trajectory ensemble.

The simulations are performed with a pulling velocity $\tilde{v}_x=2.0$.
As illustrated by the representative configurations in Fig.~\ref{fig:1}(a), this protocol progressively unfolds the hairpin through the sequential rupture of its native contacts. Increasing the native-contact strength $\epsilon_{\rm native} \equiv \epsilon$ stabilizes the folded hairpin and therefore requires increasingly stronger forces to disrupt these contacts during unfolding. As a consequence, the dissipated work increases with $\epsilon$, resulting in larger separation between the forward and reverse work distributions. This trend is illustrated in Figs.~\ref{fig:3}(a,b) for the representative interaction strengths $\epsilon/k_{\rm B}T=3$ and $12$.
For both interaction strengths shown, we find no overlap between the forward and reverse work distributions, even when using work values from $10^5$ trajectories. This makes fluctuation-theorem-based estimation of $\Delta F$ practically impossible. 
Instead, EquiNET manages to estimate the equilibrium free-energy difference directly from the forward pulling trajectories. As illustrated in Fig.~\ref{fig:1}(g), EquiNET infers the entropy production and combines it with the corresponding mean work to obtain $\Delta F^{\rm f}$, where the superscript ${\rm f}$ denotes the estimate obtained from the forward process. The resulting estimates are shown by the thick vertical dashed lines in Fig.~\ref{fig:3}(a,b), which lie well outside the range of work values realized in the sampled trajectories.

To independently validate these estimates, we need a ground-truth reference value for $\Delta F$. Since an analytical value is not available for this system owing to its strongly nonlinear dynamics and high dimensionality, we obtain a numerical reference using the Crooks fluctuation theorem applied to an additional set of forward and reverse pulling simulations performed at a slower velocity, $\tilde{v}_x=0.2$. At this slower driving rate, the reduced dissipation leads to sufficient overlap between the forward and reverse work distributions. These slow-pulling work distributions are shown in the insets of Fig.~\ref{fig:3}(a,b). Their crossing point, denoted by $W^*$ and indicated by the gray vertical line, yields $W^*=\Delta F$ according to the Crooks fluctuation theorem and thus provides an independent benchmark for the EquiNET estimates. 
The close agreement between the EquiNET estimates obtained from the strongly dissipative trajectories and $W^*$ demonstrates that EquiNET can accurately recover $\Delta F$ even in regimes where the forward and reverse work distributions exhibit no overlap.

Furthermore, EquiNET can be used even when only partial information about the system is available. This situation is directly relevant to experiments, where one typically has access only to a limited set of observables; for example, the end-to-end extension measured in optical-tweezer experiments \cite{bustamante2021optical},  fluorescence signals from a small number of labeled sites \cite{zosel2021labeling}, or distances between selected molecular markers inferred from FRET measurements \cite{ha2001single, ha2024fluorescence}.
In particular, instead of monitoring the full configuration of the hairpin, we track only the $x$ coordinates of a subset of the beads, namely the terminal bead subjected to the pulling force and the beads participating in the native contacts. 
Although these observed degrees of freedom do not fully specify the system's microscopic state, they still capture the dominant structural changes associated with hairpin unfolding. We then train EquiNET using only these reduced trajectories. Since partial observation can only decrease the inferred entropy production, the resulting estimate provides a lower bound on the true entropy production and, in turn, leads EquiNET to an upper bound on $\Delta F$. Remarkably, this upper bound remains very close to the true value, as shown by the dotted lines in Figs.~\ref{fig:3}(a,b).

Figure~\ref{fig:3}(c) summarizes the inferred free-energy differences as a function of the native-contact strength, $\epsilon/k_{\mathrm B}T$. The estimate obtained using the complete trajectory information, denoted as $\Delta F^{\rm f}$ (filled blue circles), increases approximately linearly with $\epsilon/k_{\mathrm B}T$, and is in good agreement with the independent reference value $W^*$ (filled gray diamonds) obtained from the fluctuation-theorem analysis using the slow-pulling protocol.
This linear dependence is expected because the native-contact interaction energy in Eq.~\eqref{eq:native} is directly proportional to $\epsilon_{\mathrm{native}}$.
The free-energy estimate from partial measurements, denoted by $\Delta F^{\rm f}_x$ (open blue circles), provides an upper bound on the equilibrium free-energy difference. We also report the corresponding estimates, $\Delta F^{\rm r}$ and $\Delta F^{\rm r}_x$, obtained from the reverse protocol, during which the initially extended polymer is driven back toward the folded state. Unlike the forward estimates, these provide only lower bounds on the equilibrium free-energy difference. The estimate based on the complete trajectory information, $\Delta F^{\rm r}$ (filled orange squares), lies systematically below the reference value $W^*$, while estimates obtained from partial measurement, $\Delta F^{\rm r}_x$   (open orange squares) provides an even more conservative bound.

To better understand the origin of these lower bounds in the reverse experiments, we examine the convergence properties of the inferred entropy-production rate over the course of the forward and reverse protocols, as summarized in Fig.~\ref{fig:4}.  Figures~\ref{fig:4}(a,b) show the estimated entropy-production rate for the forward protocol, for $\epsilon/k_{\rm B} T = 3$ and $12$. For each case, different curves correspond to the total number of trajectories used for inference (\(N=10^3\), \(10^4\), and \(10^{5}\)). Solid curves show the mean over six independent runs, while the shaded regions of the same color indicate \(\pm 1\) standard deviation. 
During the initial equilibration period (gray-shaded region), the entropy-production rate remains close to zero, as expected when the pulling trap is held fixed. Once the driving begins, a small but finite entropy production first develops as the hairpin is stretched away from its equilibrium configuration. This is followed by a broad, dominant peak at later times, associated with the disruption of native contacts during unfolding. The peak is substantially larger for $\epsilon/k_{\mathrm B}T=12$ than for $\epsilon/k_{\mathrm B}T=3$, consistent with the greater dissipation required to disrupt stronger native interactions. After the driving ceases, the entropy-production rate rapidly decays toward zero, indicating that the unfolded polymer relaxes sufficiently close to equilibrium during the final equilibration period.

The estimated entropy production in the reverse processes, shown in Figs.~\ref{fig:4}(c,d), displays a qualitatively different temporal profile. Following the onset of driving, the entropy-production rate increases gradually as the initially extended molecule contracts and subsequently develops an intermediate plateau-like regime. At later stages of the driving protocol, the entropy-production rate rises further as the molecule approaches the compact state and native contacts are progressively re-established. Unlike the forward process, however, the entropy-production rate does not always decay to zero within the final equilibration phase. This behavior becomes particularly evident for $\epsilon/k_{\mathrm B}T = 12$, where a finite entropy-production is seen to persist during the final equilibration period. The residual entropy production indicates that the polymer continues to relax toward the folded equilibrium state beyond the observation window and suggests that the corresponding equilibration time increases with the strength of the native interactions. Consequently, the entropy production accumulated within the time period $\tau$ has not fully converged to that associated with complete relaxation. This explains why the reverse-protocol based estimates provide only lower bounds on the actual $\Delta F$ values.

More generally, the analysis based on Figs.~\ref{fig:4}(a-d) provides a practical diagnostic for assessing whether a finite observation window of a fixed duration is sufficiently long. If the inferred entropy-production rate decays (and remains sufficiently close) to zero during the final equilibration plateau, the system can be regarded as sufficiently relaxed on the timescale of the observation. Conversely, a persistent nonzero entropy-production rate at the end of the window signals incomplete relaxation and indicates that the observation window should be extended. When extending the observation window is not feasible, the resulting free-energy estimate should instead be interpreted as a bound.

Finally, to quantify the data requirements of EquiNET, we performed a parameter sweep over the number of pulling trajectories used for inference. 
Figures~\ref{fig:4}(e,f) show the fraction of the total entropy production recovered by the inference as a function of the total number of trajectories, for $\epsilon/k_{\mathrm B}T=3$ and $12$. 
Here, the ground-truth total entropy production is determined independently from the work statistics as $ \langle \Delta S_{\mathrm{tot}}^{\mathrm{true}}\rangle = \langle W_{\rm F}\rangle-W^*$ for the forward process and $\langle \Delta S_{\mathrm{tot}}^{\mathrm{true}}\rangle = \langle W_{\rm R}\rangle+W^*$ for the reverse process, where $W^*$ is the estimate of $\Delta F$ obtained from the slow protocol. The horizontal line at unity therefore corresponds to complete recovery of the true entropy production. For both protocols, the inferred fraction approaches a well-defined large-sample value as the dataset size increases. The forward inference converges particularly rapidly, recovering nearly all of the total entropy production, whereas the reverse inference converges more slowly and captures a smaller fraction, particularly for the stronger interaction strength. We note that, in the forward process, approximately \(10^3\) trajectories are sufficient to recover a large fraction of the true entropy production. This means, EquiNET can extract the relevant irreversible signal from comparatively modest datasets.



In summary, we have introduced EquiNET, an inference technique for estimating equilibrium free energy differences from non-equilibrium trajectory fluctuations by directly inferring the corresponding total entropy production. In contrast to conventional approaches based on fluctuation relations, which rely on rare events and become increasingly data-inefficient in strongly dissipative regimes, our method relies on inferring the average entropy production and using that information to correct for dissipated work to recover $\Delta F$. Since this approach only relies on lower order statistics (mean and variance) of trajectory observables, it is found to converge faster than the fluctuation theorem. 

In its most general form, EquiNET is implemented using a physics-informed machine learning framework based on a time-conditioned E(3)-equivariant graph neural network. This provides a scalable route for analyzing high-dimensional systems, such as complex biomolecules or optical matter assemblies, whenever the underlying interaction network is known \emph{a priori}. As formulated, EquiNET requires measurements of all degrees of freedom necessary to render the observed dynamics Markovian to provide an exact estimate of $\Delta F$, which is not always experimentally accessible. Nevertheless, the framework naturally extends to partial observations in which only a subset of these degrees of freedom are measured. Although this inevitably yields lower bounds on the entropy production and corresponding upper bounds on the equilibrium free-energy difference, these bounds can be significantly tightened by judiciously selecting the observed degrees of freedom. For instance, in biomolecular pulling experiments involving multi-domain proteins or nucleic acids, one often measures only the end-to-end extension while the internal conformational dynamics remain hidden. Our results suggest that incorporating additional observables, for example, through site-specific labeling or super-resolution fluorescence microscopy \cite{roy2008practical, ha2001single}, could substantially tighten the inferred bounds on entropy production and, consequently, on the equilibrium free-energy difference. More generally, further improvements may be achieved by combining complementary information from multiple independent labeling experiments \cite{hellenkamp2017multidomain}. 

Finally, although we focus here on overdamped diffusive processes, extending EquiNET beyond this setting presents several interesting directions for future work. Recent trajectory-based approaches for underdamped dynamics and systems with odd-parity variables provide promising routes toward inference in full phase space \cite{Kim2022,BoffiVandenEijnden2025}. A complementary route for extending the framework to discrete space Markov jump processes can be achieved using the estimator obtained in Refs.~\cite{NEEP, otsubo2022estimating}. Exploring these connections may considerably broaden the range of nonequilibrium systems for which equilibrium free-energy differences can be inferred directly from trajectory fluctuations. 

\section*{Data and Code Availability}
The code used to generate the data, train the neural networks, and reproduce all figures and analyses in this study is available at \href{https://github.com/softmatterlab/EquiNET}{github.com/softmatterlab/EquiNET}
and \href{https://doi.org/10.5281/zenodo.22083289}{doi.org/10.5281/zenodo.22083289}.

\section*{Acknowledgements}
SKM acknowledges the Knut and Alice Wallenberg Foundation for financial support through Grant No. KAW 2023.0149. Parts of the computations were enabled by resources provided by the National Academic Infrastructure for Supercomputing in Sweden (NAISS), partially funded by the Swedish Research Council through grant agreement no. 2022-06725. 
GV acknowledges support from
the Horizon Europe ERC Consolidator Grant MAPEI
(grant number 101001267), from the Knut and Alice
Wallenberg Foundation (grant number 2019.0079), and
from the Göran Gustafsson Foundation for Research in
Natural Sciences and Medicine.


%

\appendix

\section{Inference Algorithm}

In Appendix~A, we discuss the construction of the neural network and the inference algorithm in detail.
\subsection{Neural Network design}
In this study, we use a time-conditioned E(3)-equivariant graph neural network (EGNN \cite{satorras2021n}) to parameterize the unknown thermodynamic force field, $\mathcal{\bm F}(\mathbf{x},t)$. EGNN is particularly well suited for molecular systems, such as the polymer hairpin example we considered here, because they operate directly on molecular coordinates while preserving the translational, rotational, and reflection symmetries of three-dimensional space. Furthermore, representing a molecule as a graph enables the network to naturally exploit its underlying topology. 

To implement this architecture, each configuration of the polymer hairpin is first represented as a directed graph whose nodes correspond to polymer beads. The graph connectivity is constructed to encode the dominant structural interactions in the system. Specifically, bidirectional edges are introduced between nearest-neighbor beads along the polymer backbone, next-nearest-neighbor beads, and the predefined native-contact pairs. 

Each node is associated with its Cartesian coordinates,
\begin{equation}
\mathbf{x}_i^{(0)}\equiv\mathbf{x}_i^{t}\in\mathbb{R}^3,
\end{equation}
and an initial node feature,
\begin{equation}
\mathbf{h}_i^{(0)}
=
\mathbf{e}_i+\mathbf{e}_t,
\end{equation}
where $\mathbf{e}_i \in\mathbb{R}^h$ denotes a learnable bead embedding and $\mathbf{e}_t \in\mathbb{R}^h$ is a global embedding of time, $t$ which is set to be the same for all the beads. The temporal embedding is constructed from a set of $K$ Gaussian basis functions with learnable centers $\{\mu_k\}$ and widths $\{\sigma_k\}$,
\begin{equation}
\phi_k(t)
=
\exp\!\left[
-\frac{1}{2}
\left(
\frac{t-\mu_k}{\sigma_k}
\right)^2
\right],
\end{equation}
where each basis function is localized around its  center $\mu_k$ with characteristic width $\sigma_k$, both of which are treated as learnable parameters. Collectively, these basis functions provide a smooth nonlinear representation of time $t$, allowing the network to capture time-dependent changes in the optimal coefficient field. 

The resulting $K$-dimensional time representation,
\begin{equation}
\bm{\phi}(t)=
\left[
\phi_1(t),
\phi_2(t),
\ldots,
\phi_K(t)
\right]^{\!\top}
\in\mathbb{R}^K,
\end{equation}
is subsequently projected into the hidden feature space $\mathbf{e}_t \in \mathbb{R}^h$ through a learnable linear transformation. 

The graph is processed through a stack of $L$ EGNN layers. At each layer, edge messages are constructed from the node embeddings of the connected beads together with their squared Euclidean separation. For an edge connecting nodes $i$ and $j$, the relative displacement vector
\begin{equation}
\mathbf{r}_{ij}^{(l)}
=
\mathbf{x}_i^{(l)}
-
\mathbf{x}_j^{(l)}
\end{equation}
and the squared distance
\begin{equation}
r_{ij}^{2,(l)}
=
\left\|
\mathbf{x}_i^{(l)}
-
\mathbf{x}_j^{(l)}
\right\|^2
\end{equation}
are computed. These quantities are concatenated with the sender and receiver node embeddings and passed through the edge-update multilayer perceptron,
\begin{equation}
\mathbf{m}_{ij}^{(l)}
=
\phi_e
\!\left(
\mathbf{h}_i^{(l)},
\mathbf{h}_j^{(l)},
r_{ij}^{2,(l)}
\right),
\end{equation}
to obtain the edge messages.

The edge messages serve two purposes. First, they are transformed by a coordinate multilayer perceptron,
\begin{equation}
w_{ij}^{(l)}
=
\phi_x
\!\left(
\mathbf{m}_{ij}^{(l)}
\right),
\end{equation}
which produces scalar coordinate-update weights. These weights determine the contribution of each neighboring bead to the coordinate update,
\begin{equation}
\mathbf{x}_i^{(l+1)}
=
\mathbf{x}_i^{(l)}
+
\sum_{j}
w_{ij}^{(l)}
\left(
\mathbf{x}_j^{(l)}
-
\mathbf{x}_i^{(l)}
\right),
\end{equation}
Since the coordinate update is expressed as a weighted sum of relative
displacement vectors multiplied by learned scalar coefficients, it
transforms equivariantly under translations, rotations, and reflections.
(This Euclidean equivariance gives rise to the name
\emph{E($3$)-equivariant graph neural network} (EGNN) in three dimensions.)

Second, the incoming edge messages are aggregated at each node,
\begin{equation}
\mathbf{m}_i^{(l)}
=
\sum_j
\mathbf{m}_{ij}^{(l)},
\end{equation}
and concatenated with the current node embedding. The resulting vector is passed through the node-update multilayer perceptron,
\begin{equation}
\Delta\mathbf{h}_i^{(l)}
=
\phi_h
\!\left(
\mathbf{h}_i^{(l)},
\mathbf{m}_i^{(l)}
\right),
\end{equation}
to obtain the node-feature update,
\begin{equation}
\mathbf{h}_i^{(l+1)}
=
\mathbf{h}_i^{(l)}
+
\Delta\mathbf{h}_i^{(l)}.
\end{equation}

The updated coordinates and node embeddings are propagated to the subsequent EGNN layer, where the edge messages are recomputed from the current graph representation. 
Following the final EGNN layer, the node embeddings are passed independently through a two-layer readout network,
\begin{equation}
\mathbf{d}_i(\mathbf{x},t)
=
\phi_d
\!\left(
\mathbf{h}_i^{(L)}
\right),
\end{equation}
which predicts a three-dimensional vector for every bead. Collectively, these vectors define the bead-wise coefficient field,
\begin{equation}
\mathbf{d}(\mathbf{x},t)
=
\left\{
\mathbf{d}_i(\mathbf{x},t)
\right\}_{i=1}^{N},
\end{equation}
that is used in the variational objective in Eq.~\eqref{eq:tur}, to infer the instantaneous entropy production rate, $\sigma(t)$. Separate EGNNs with identical architectures are trained independently for the forward and reverse trajectory ensembles.

\subsection{Algorithm}
To implement the inference algorithm, we use $N_{\mathrm{train}} = N/2$ trajectories as the training set, and  another $N_{\mathrm{test}} = N/2$ trajectories as the independent test set. 
For each trajectory, consecutive stored trajectory points are converted into midpoint configurations,
\begin{equation}
\bar{\mathbf{x}}_{r,n}
=
\frac{\mathbf{x}_{r,n}+\mathbf{x}_{r,n+1}}{2},
\end{equation}
and finite displacements,
\begin{equation}
\Delta\mathbf{x}_{r,n}
=
\mathbf{x}_{r,n+1}
-
\mathbf{x}_{r,n},
\end{equation}
where $r$ labels trajectories and $n=0,\ldots,L-1$ labels trajectory points. 

Separate EGNNs with identical architectures are trained for the forward and reverse trajectory ensembles. At each optimization step, a minibatch of $B_{\mathrm{traj}}$
training trajectories is sampled uniformly without replacement. Independently, a minibatch of $B_t$
time indices is sampled uniformly without replacement. For every sampled trajectory $r$ and time point $t_n$, the EGNN predicts the bead-wise vector field
\begin{equation}
\mathbf d_{\boldsymbol\theta}
(\bar{\mathbf x}_{r,n},t_n),
\end{equation}
from which the scalar current is computed as
\begin{equation}
J_{r,n}
=
\sum_{i=1}^{N}
\mathbf d_{\boldsymbol\theta,i}
(\bar{\mathbf x}_{r,n},t_n)
\cdot
\Delta\mathbf x_{r,n,i}.
\end{equation}

For every sampled time point, the minibatch mean
\begin{equation}
\langle J\rangle_n
=
\frac{1}{B_{\mathrm{traj}}}
\sum_{r=1}^{B_{\mathrm{traj}}}
J_{r,n},
\end{equation}
and the unbiased sample variance
\begin{equation}
\operatorname{Var}(J)_n
=
\frac{1}{B_{\mathrm{traj}}-1}
\sum_{r=1}^{B_{\mathrm{traj}}}
\left(
J_{r,n}
-
\langle J\rangle_n
\right)^2
\end{equation}
are computed over the trajectory minibatch.

The network parameters are optimized by minimizing the negative variational objective,
\begin{equation}
\mathcal L(\boldsymbol\theta)
=
-
\sum_{n\in\mathcal T}
\frac{
2\langle J\rangle_n^2
}{
\Delta t_{\mathrm{inf}}
\left[
\operatorname{Var}(J)_n
\right]
},
\end{equation}
where $\mathcal T$ denotes the sampled time minibatch. Adam optimization is employed together with gradient clipping and a cosine-annealing learning-rate schedule.

Following training, the network is evaluated exclusively on the independent held-out test trajectories. Midpoint configurations and displacements are reconstructed from the test trajectories, and the corresponding currents,
\begin{equation}
J_{r,n}^{\mathrm{test}}
=
\sum_{i=1}^{N}
\mathbf d_{\boldsymbol\theta,i}
(\bar{\mathbf x}_{r,n}^{\mathrm{test}},t_n)
\cdot
\Delta\mathbf x_{r,n,i}^{\mathrm{test}},
\end{equation}
are computed without updating the network parameters. 

The instantaneous entropy production rate is finally estimated as
\begin{equation}
\widehat{\sigma}(t_n)
=
\frac{
2
\left(
\langle J\rangle_n^{\mathrm{test}}
\right)^2
}{
\Delta t_{\mathrm{inf}}
\left[
\operatorname{Var}(J)_n^{\mathrm{test}}
\right]
},
\end{equation}
and the total entropy production is obtained by numerical integration,
\begin{equation}
\widehat{\Delta S}_{\mathrm{tot}}
=
\Delta t_{\mathrm{inf}}
\sum_{n=0}^{L-1}
\widehat{\sigma}(t_n).
\end{equation}

\begin{table}[t]
\centering
\caption{Hyperparameters used for training the time-conditioned EGNN.}
\label{tab:hyperparameters}
\begin{tabular}{ll}
\hline
\textbf{Hyperparameter} & \textbf{Value} \\
\hline
Optimizer & Adam \\
Initial learning rate & $10^{-2}$ \\
Learning-rate schedule & Cosine annealing \\
Minimum learning rate & $10^{-5}$ \\
Training epochs & $10\,000$ \\
Gradient clipping norm & $5.0$ \\
Trajectory minibatch size & $\min(5000,N_{\mathrm{train}})$ \\
Time minibatch size & $\min(10,L)$ \\
Number of Gaussian basis functions ($K$) & $30$ \\
Hidden embedding dimension ($h$) & $32$ \\
Number of EGNN layers & $2$ \\
Edge MLP $\phi_e$ & $(2h+1)\rightarrow h\rightarrow h$ \\
Coordinate MLP $\phi_x$ & $h\rightarrow h\rightarrow 1$ \\
Node MLP $\phi_h$ & $(2h)\rightarrow h\rightarrow h$ \\
Readout MLP $\phi_d$ & $h\rightarrow h\rightarrow d_{\mathrm{coord}}$ \\
Activation function & SiLU \\
Variance regularization ($\varepsilon$) & $10^{-6}$ \\
Coordinate dimension (full / CG) & $3$ / $1$ \\
\hline
\end{tabular}
\end{table}

\subsection{Coarse-grained representation and inference}

For the coarse-grained model, the full system is simulated, and coarse-graining is subsequently performed by retaining only the x-coordinates of a selected subset of beads. The selected set contains the beads participating in native contacts together with the terminal bead. Accordingly, the full molecular representation,
\begin{equation}
\mathbf{x}\in\mathbb{R}^{N\times 3},
\end{equation}
is replaced by the reduced representation
\begin{equation}
\mathbf{x}_{\mathrm{CG}}
\in
\mathbb{R}^{N_{\mathrm{CG}}\times 1},
\end{equation}
where $N_{\mathrm{CG}}$ denotes the number of retained beads.

The coarse-grained graph is obtained by restricting the original molecular graph to the selected beads. Backbone, next-nearest-neighbor, and native-contact edges are retained only when both of their endpoints belong to the selected bead set. Each retained physical connection is represented by two directed edges, allowing bidirectional message passing. Thus, the coarse-grained topology is the subgraph induced by the selected nodes.

The same time-conditioned EGNN architecture is used for the full and coarse-grained representations. The main architectural difference is that the coordinate dimension is reduced from three to one. Consequently, the relative coordinate entering the EGNN layer is the scalar difference $
r_{ij}^{(l)}
=
x_i^{(l)}-x_j^{(l)}$,
and the corresponding invariant edge feature is $
\left(r_{ij}^{(l)}\right)^2
=
\left(x_i^{(l)}-x_j^{(l)}\right)^2$. The output of the readout network is likewise reduced from a three-dimensional vector, $\mathbf{d}_i(\mathbf{x},t)\in\mathbb{R}^3$ to a scalar coefficient for each retained bead,
$d_i(\mathbf{x}_{\mathrm{CG}},t)\in\mathbb{R}$. The training and inference procedures are otherwise unchanged. 

\section{Dynamics in a one dimensional, time-dependent potential}

The overdamped Langevin dynamics are driven by a time-dependent potential that smoothly transforms an initial harmonic well into an asymmetric double-well potential and subsequently into a final harmonic well. The double-well potential is defined as

\begin{align}
U_{\rm DW}(x)=\left(x^2-1\right)^2+ax^3,
\end{align}

where \(a\) controls the asymmetry. The stationary points satisfy

\begin{align}
\frac{dU_{\rm DW}}{dx}
=
4x(x^2-1)+3ax^2
=
x(4x^2+3ax-4)=0.
\end{align}

The two minima are therefore located at

\begin{align}
x_{L,R}
=
\frac{-3a\mp\sqrt{9a^2+64}}{8},
\end{align}

with corresponding energies

\begin{align}
U_{L,R}=U_{\rm DW}(x_{L,R}),
\end{align}

and local curvatures

\begin{align}
k_{L,R}
=
\left.
\frac{d^2U_{\rm DW}}{dx^2}
\right|_{x=x_{L,R}},
\end{align}

where

\begin{align}
\frac{d^2U_{\rm DW}}{dx^2}
=
12x^2+6ax-4.
\end{align}

The initial and final harmonic potentials are constructed to match the energy and curvature of the corresponding minima,

\begin{align}
U_L(x)
=
U_L
+\frac{k_L}{2}(x-x_L)^2,
\end{align}

and

\begin{align}
U_R(x)
=
U_R
+\frac{k_R}{2}(x-x_R)^2.
\end{align}

The complete time-dependent potential is

\begin{align}
U(x,t)
=
\alpha(t)U_L(x)
+
\beta(t)U_{\rm DW}(x)
+
\gamma(t)U_R(x),
\end{align}

where

\begin{align}
\alpha(t)+\beta(t)+\gamma(t)=1.
\end{align}

The protocol consists of an initial equilibration period
\(t_{\rm eq,i}\), a driving period \(t_{\rm p}\), and a final
equilibration period \(t_{\rm eq,f}\), with total duration

\begin{align}
\tau
=
t_{\rm eq,i}
+
t_{\rm p}
+
t_{\rm eq,f}.
\end{align}

During the driving period the interpolation weights evolve according to
the cubic smoothstep function

\begin{align}
S(u)=u^2(3-2u),
\end{align}

so that the potential evolves continuously according to

\begin{align}
U_L
\rightarrow
U_{\rm DW}
\rightarrow
U_R.
\end{align}

The interpolation weights are

\begin{align}
\alpha(t)=
\scalebox{0.8}{$\begin{cases}
1,
&
t\le t_{\rm eq,i},
\\[2mm]
1-
S\!\left(
\dfrac{2(t-t_{\rm eq,i})}{t_{\rm p}}
\right),
&
t_{\rm eq,i}
<
t
<
t_{\rm eq,i}
+\dfrac{t_{\rm p}}{2},
\\[3mm]
0,
&
t\ge
t_{\rm eq,i}
+\dfrac{t_{\rm p}}{2},
\end{cases}$}
\end{align}

\begin{align}
\beta(t)=
\scalebox{0.8}{$\begin{cases}
0,
&
t\le t_{\rm eq,i},
\\[2mm]
S\!\left(
\dfrac{2(t-t_{\rm eq,i})}{t_{\rm p}}
\right),
&
t_{\rm eq,i}
<
t
<
t_{\rm eq,i}
+\dfrac{t_{\rm p}}{2},
\\[3mm]
1-
S\!\left(
\dfrac{2(t-t_{\rm eq,i})}{t_{\rm p}}-1
\right),
&
t_{\rm eq,i}
+\dfrac{t_{\rm p}}{2}
<
t
<
t_{\rm eq,i}
+t_{\rm p},
\\[3mm]
0,
&
t\ge
t_{\rm eq,i}
+t_{\rm p},
\end{cases}$}
\end{align}

\begin{align}
\gamma(t)=
\scalebox{0.8}{$\begin{cases}
0,
&
t<
t_{\rm eq,i}
+\dfrac{t_{\rm p}}{2},
\\[3mm]
S\!\left(
\dfrac{2(t-t_{\rm eq,i})}{t_{\rm p}}-1
\right),
&
t_{\rm eq,i}
+\dfrac{t_{\rm p}}{2}
<
t
<
t_{\rm eq,i}
+t_{\rm p},
\\[3mm]
1,
&
t\ge
t_{\rm eq,i}
+t_{\rm p}.
\end{cases}$}
\end{align}

The equilibrium free-energy difference between the initial and final
harmonic states can be evaluated analytically. Using the Einstein
relation

\begin{align}
k_BT=\frac{D}{\mu},
\end{align}

the partition function of a harmonic well is

\begin{align}
Z
=
e^{-\beta U_0}
\sqrt{\frac{2\pi}{\beta k}},
\end{align}

which gives the free energy

\begin{align}
F
=
U_0
-
\frac{k_BT}{2}
\ln\left(
\frac{2\pi}{\beta k}
\right).
\end{align}

The equilibrium free-energy difference therefore becomes

\begin{align}
\Delta F
=
F_R-F_L
=
(U_R-U_L)
+
\frac{k_BT}{2}
\ln\left(
\frac{k_R}{k_L}
\right),
\end{align}

which provides the exact equilibrium reference used to validate the
nonequilibrium free-energy estimators presented in the main text.

The simulations use the parameter values
\(
a=-1,\;
\mu=1,\;
D=0.05,\;
t_{\rm eq,i}=0.9,\;
t_{\rm p}=1.2,\;
t_{\rm eq,f}=0.9.
\)
The total protocol duration is
\(
\tau=t_{\rm eq,i}+t_{\rm p}+t_{\rm eq,f}=3.
\)
The overdamped Langevin equation is integrated using the Euler--Maruyama scheme with a time-step of $\Delta t = 10^{-3}s$, with an ensemble of \(N=10^4\) trajectories. The corresponding free energy difference is $\Delta F = -48.36\; k_{\rm B} T$.
\section{Simulation parameters for the polymer hairpin model}
\label{app:hairpin_parameters}

The polymer consists of $n=13$ beads in three spatial dimensions connected by harmonic bonds and stabilized by three native contacts. The first bead is attached to a fixed harmonic anchor, while the final bead is attached to a harmonic trap that is translated during the pulling protocol. The model parameters are summarized in Table~\ref{tab:hairpin_parameters}.

\begin{table}[h]
\centering
\caption{Simulation parameters for the polymer hairpin model.}
\label{tab:hairpin_parameters}
\begin{tabular}{lc}
\hline
Parameter & Value\\
\hline
Number of beads, $n$ & 13\\
Spatial dimension & 3\\
Thermal energy, $k_{\rm B}T$ & 1\\
Friction coefficient, $\gamma$ & 1\\
Integration time step, $\Delta t$ & $10^{-5}$\\
Data storing interval, $\Delta t_{\rm inf}$ & $5 \times 10^{-2} $\\
Bond spring constant, $k_{\rm bond}$ & 500\\
Bending stiffness, $k_{\rm bend}$ & 20\\
Anchor spring constant, $k_{\rm anchor}$ & 1000\\
Pulling spring constant, $k_{\rm pull}$ & 10\\
Native contact strength, $\epsilon_{\rm native}$ & 12\\
Repulsive interaction strength, $\epsilon_{\rm rep}$ & 1\\
Repulsive interaction length scale, $\sigma_{\rm rep}$ & 0.8\\
Pulling speed, $v_x$ & 2\\
Pulling speed for slow protocol, $v_x$ & 0.2\\
\hline
\end{tabular}
\end{table}

\begin{table*}[h]
\centering
\caption{
\textbf{Numerical free-energy estimates for the polymer hairpin model.}
The Crooks crossing $W^\ast$ is obtained from the slow-driving work distributions.
The forward and reverse free-energy estimates are computed from the corresponding mean work and inferred entropy production, with the subscript $x$ denoting the coarse-grained estimate.
Uncertainties in the free-energy estimates are obtained by propagating the uncertainties in the mean work and inferred entropy production. Reported uncertainties correspond to the standard deviation across six runs, which use different combination of trajectory realizations.
}
\vspace{2mm}
\label{tab:free_energy_estimates}
\setlength{\tabcolsep}{10pt}
\begin{tabular}{cccccc}
\hline
$\epsilon/k_{\rm B}T$
& $W^\ast/k_{\rm B}T$
& $\Delta F^{\rm f}/k_{\rm B}T$
& $\Delta F^{\rm f}_{x}/k_{\rm B}T$
& $\Delta F^{\rm r}/k_{\rm B}T$
& $\Delta F^{\rm r}_{x}/k_{\rm B}T$ \\
\hline

3
& $8.983 \pm 0.052$
& $7.637 \pm 0.099$
& $33.220 \pm 0.075$
& $0.547 \pm 0.061$
& $-11.724 \pm 0.050$ \\

6
& $14.570 \pm 0.100$
& $15.526 \pm 0.095$
& $41.591 \pm 0.106$
& $1.355 \pm 0.118$
& $-11.014 \pm 0.093$ \\

9
& $22.518 \pm 0.038$
& $24.119 \pm 0.122$
& $50.861 \pm 0.102$
& $3.537 \pm 0.073$
& $-9.568 \pm 0.118$ \\

12
& $30.698 \pm 0.070$
& $32.489 \pm 0.114$
& $59.455 \pm 0.150$
& $8.349 \pm 0.110$
& $-6.551 \pm 0.187$ \\

\hline
\end{tabular}

\end{table*}

\section{Free-energy estimates}
\label{app:free_energy_estimates}

In Table.~\ref{tab:free_energy_estimates}, we report the numerical values of the free-energy estimates shown in Fig.~\ref{fig:3}.

\end{document}